\documentclass[preprint,aps,floatfix,superscriptaddress,prx,showpacs]{revtex4}
\usepackage{amsmath, amsthm, amssymb, graphicx}
\usepackage{color}
\usepackage{multirow}
\usepackage{ulem}

\input epsf.sty
\begin{document}

\title{Foraging patterns in online searches}
\author{Xiangwen Wang}
\affiliation{Department of Physics, Virginia Tech, Blacksburg, VA 24061-0435, USA}
\affiliation{Center for Soft Matter and Biological Physics, Virginia Tech, Blacksburg, VA 24061-0435, USA}
\affiliation{Department of Statistics, Virginia Tech, Blacksburg, VA 24061-0439, USA}
\author{Michel Pleimling}
\affiliation{Department of Physics, Virginia Tech, Blacksburg, VA 24061-0435, USA}
\affiliation{Center for Soft Matter and Biological Physics, Virginia Tech, Blacksburg, VA 24061-0435, USA}
\affiliation{Academy of Integrated Science, Virginia Tech, Blacksburg, VA 24061-0405, USA}
\date{\today}

\begin{abstract}
Nowadays online searches are undeniably the most common form of information gathering, as witnessed by
billions of clicks generated each day on search engines. In this work
we describe online searches as foraging processes that take place on the semi-infinite 
line. Using a variety of quantities like probability distributions and 
complementary cumulative distribution functions of step-length and
waiting time as well as mean square displacements and entropies, we analyze three different click-through logs 
that contain the detailed information of millions of queries submitted to search engines.
Notable differences between the different logs reveal an increased efficiency
of the search engines. In the language of foraging, the newer logs indicate that online searches
overwhelmingly yield local searches (i.e. on one page of links provided by the search engines),
whereas for the older logs the foraging processes are a combination of local searches
and relocation phases that are power law distributed. 
Our investigation of
click logs of search engines therefore highlights the presence of intermittent search processes
(where phases of local explorations are separated by power law distributed relocation jumps)
in online searches.
It follows that good search engines enable the users to find the information they
are looking for through a local exploration of a single page with search results, whereas for poor 
search engines users are often forced to do a broader exploration of different pages.

\end{abstract}


\maketitle

\section{Introduction}

An increasingly large part of our day is devoted to online
activities. It is therefore not surprising that in recent years online mobility patterns have
emerged as a new interdisciplinary research area. Much attention has been given to scaling 
and non-Markovian features of web browsing \cite{Huberman1998,Fagin2001,Chmiel2009,
Chierichetti2012,Zhao2014,Zhao2015}, to the features of mobility in online games \cite{Szell2012,Sinatra2014}
as well as to emerging scaling properties in e-commerce \cite{Zhao2013}.

Each day, tens of billions of clicks are generated on search engines. Understanding human
online search click-through behavior can therefore be of central importance to improve ranking algorithms,
rearrange page layout for search engines, and reduce advertisement spending for enterprises.
Click-through data, which are extracted from the click logs of search engines, contain information
on the links clicked by a user as a result of a query submitted to a search engine. These
data have been exploited in a variety of studies that aimed at optimizing web searches and at
improving retrieval quality \cite{Joachims2002,Xue2004,Craswell2007,Radlinski2008}.

Our daily experience with web searches shows that a fully deterministic search
strategy usually does not optimize the search outcome. Instead, there is often some degree of
randomness involved in choosing the links to click on among those provided by a search engine.
It is therefore tempting to investigate web searches from the point of view of random search strategies.

Studies of search strategies \cite{Viswanathan2011} in animal foraging
\cite{Edwards2007,Sims2008,Humphries2010,deJaeger2011,Humphries2012,Schultheiss2015} hint at
intriguing connections between movement patterns and availability of preys.
Roughly speaking, one can distinguish between two different cases.
If the prey is abundant, then the predator tends to perform a random
walk and only explores a rather restricted territory. On the other hand,
if the prey is scarce, evidence has been found that the pattern changes to a L\'{e}vy walk (or, alternatively,
to an intermittent search process that includes L\'{e}vy movements \cite{Lon08,Ben11}) that allows
to cover much larger areas/volumes and to
optimize the search efficiency when resources are
sparsely distributed. It has been claimed that L\'{e}vy movement patterns also show up in human foraging
\cite{Brown2007,Raichlen2014} as well as in the migration of bacteria \cite{Korobkova2004,Ariel2015}
and T cells \cite{Harris2012}. 
Of course, the simple relationship between displacement pattern and availability
of preys only prevails
on large scales, and a much more complex and subtle picture emerges when going beyond
such a coarse-grained description \cite{Edwards2011,Kolzsch2012,Pyke2015,Reynolds2015}. 

Analyzing extensive
click-through data sets from different search engines collected in different years,
we consider in this paper online searches as foraging processes on a straight semi-infinite line and
find a transition in the search patterns from a behavior that includes long-range relocations
to a purely local Brownian-type motion with increasing efficiency of the search engines.
A more detailed analysis reveals a behavior
that is more complex than simple L\'{e}vy flights or Brownian motions.

In the next Section we describe the click-through data as trajectories on a semi-infinite line.
Section III analyzes these data as foraging processes on the semi-infinite line through
the study of numerous quantities, ranging from probability distributions and
complementary cumulative distribution functions of displacement and
waiting time to mean square displacements and entropies.
Our analysis indicates that the character of online searches changes with increasing 
efficiency of search engines, shifting from processes that include power law distributions
to a Brownian-type behavior.

\section{Click-through data sets}
Our study of human online search patterns is based on three click-through data sets
collected by different commercial search engine providers. The sets 
Sogou-08 and Sogou-11 were collected 
on the Chinese search engine Sogou in 2008 and 2011 respectively, whereas the set Yahoo-10 was collected on 
Yahoo in 2010. For a detailed description 
of the data sets and of the data preparation we refer the reader to Appendix A.

\begin{figure} [h]
\includegraphics[width=0.70\columnwidth,clip=true]{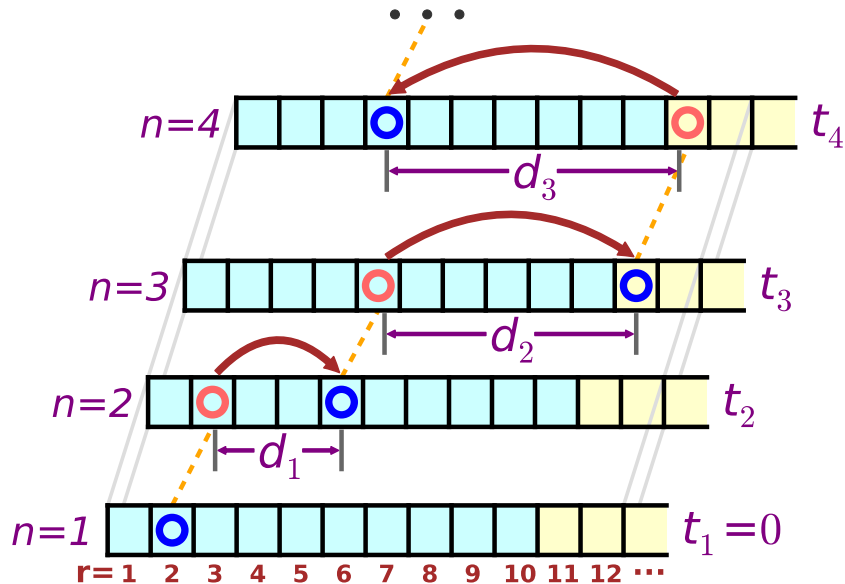}
\caption{\label{fig1} 
(Color online) Illustration of the terms used in this study: clicking order $n$, rank $r$, time of $n$th
click $t_n$, and step-length $d_n$ between successive clicks on results provided by the search engine.
}
\end{figure}

As explained in Appendix A and illustrated in Fig. \ref{fig1},
for  each query submitted to one of the search engines
we assign to every click on a search result a pair of ``space"-``time" coordinates
where the ``time" is the time in seconds passed since the first click (i.e. 
for the first click, $t_1=0$), whereas the ``space" coordinate is the rank of the
search result, i.e. its position when treating all search results as points on a 
semi-infinite line, where the top result is assigned the rank $r =1$. The $n$th click
is therefore characterized by the pair $(t_n,r_n)$. Subsequent clicks 
for a given search then correspond to subsequent steps along the semi-infinite line
where the rank difference $\Delta r_n = \left| r_{n+1}-r_n \right|$ is
the step-length $d_n = \Delta r_n$. The sign of $r_{n+1}-r_n$ provides us with the
direction of the steps. As we will see later, the data show a strong bias in the forward 
direction. 
In a similar way we define as waiting time 
$\tau_n$ the time interval between two successive clicks \cite{footnote1}: $\tau_n = t_{n+1} - t_n$.


Usually a query ends when the user finds the relevant information. In the language of foraging
the relevant information is therefore the resource. We know from our
own experience that users occasionally terminate a search early once they are convinced that the
search term is unlikely to yield the expected result. The limitations of the data bases used
in this study do not allow to identify these instances. We therefore do not try to distinguish between
these cases and treat all queries in the same way, with the 
resource being located at the site $r = r_f$ that corresponds to the rank of the last
search result clicked by the user. Of interest is also the number of clicks (steps) needed to reach the
resource. As we discuss below, the resource location $r_f$ and the clicking number $N_c$ provide simple ways of evaluating the efficiency
of a search engine.
Both $r_f$ and $N_c$ would presumably change slightly if we would be able to
identify those searches that resulted in the user finding the information they
were looking for.

In the following we focus on probability distributions and on complementary cumulative distribution functions
(CCDF(x) is the probability $P(X>x)$ that the random variable $X$ has a value larger than $x$) in order to analyze the motion patterns 
emerging from online searches. We take a population level approach and base our analysis on population-averaged
distributions. As we are dealing with millions of queries for every search engine, see Table I, we expect that
distribution functions provide a very reliable characterization of the foraging patterns in online
searches.

\section{Model selection and parameter estimation}
In the following we are modeling the distributions of various quantities in the large-value limit through a variety of models
and select the best model using the Akaike information criterion (AIC). As the quantities derived
from the click-through data sets take on positive integers only, we are considering only discrete versions
of the models. 

As a power-law model we use the discrete power-law (DPL) distribution
\begin{equation} \label{discrete-power-law}
\displaystyle P(k) = \frac{k^{-\alpha}}{\zeta(\alpha, k_{\text{min}})} \sim k^{-\alpha}, \quad k \geq k_{\text{min}},\ \alpha > 1,
\end{equation}
where the normalizing factor $\displaystyle \zeta(\alpha, k_{\text{min}})=\sum\limits_{m=k_\text{min}}^\infty m^{-\alpha}$ is the incomplete $\zeta$-function \cite{Newman2005}.
For the exponential model 
we use the ``shifted'' geometric (SG) distribution
\begin{eqnarray} 
\displaystyle P(k) &=& p(1-p)^{k-k_{\text{min}}}, \quad k \geq k_{\text{min}},\ 1\geq p >0,\label{shifted-geometric}\\
&=& \left( 1-e^{-\lambda} \right) e^{-\lambda(k-k_\text{min})} \sim e^{-\lambda k}, \quad \lambda > 0,
\end{eqnarray}
where $\lambda = -\ln (1-p)$.

We also included in the model selection the power law with exponential cut-off (PEC) model:
\begin{equation} \label{power-law-with-exponential-cutoff}
\displaystyle P(k) = \frac{1}{\displaystyle Li_\alpha(e^{-\lambda})-\sum_{i=1}^{k_{\text{min}-1}}i^{-\alpha} e^{-\lambda i}} k^{-\alpha} e^{-\lambda k} \sim k^{-\alpha} e^{-\lambda k}, \quad k \geq k_{\text{min}},\ \lambda > 0,\ \alpha > 0,
\end{equation}
where $\displaystyle Li_\alpha(z)$ is the polylogarithm function, 
the discrete log-normal model (DLN) \cite{Chakraborty2015}:
\begin{equation} \label{discrete-lognormal}
\displaystyle P(k) = \frac{\displaystyle \Phi \left(\frac{\ln(k+1)-\mu}{\sigma}\right)-\Phi\left(\frac{\ln(k)-\mu}{\sigma} \right)}{\displaystyle 1-\Phi \left(\frac{\ln(k_\text{min})-\mu}{\sigma} \right)}, \quad k \geq k_{\text{min}},\ \sigma > 0,
\end{equation}
where $\Phi(\cdot)$ is the standard normal cumulative distribution function, the Yule-Simon (YS) distribution \cite{Simon1955,Clauset2009}
\begin{equation}\label{yule-simon}
\displaystyle P(k) = (\alpha -1) \frac{\Gamma\left(k_\text{min}+\alpha +1\right)}{\Gamma\left( k_\text{min} \right)} \frac{\Gamma(k)}{\Gamma(k+\alpha)}, \quad k \geq k_{\text{min}},\ \alpha > 1,
\end{equation}
which for $k$ sufficiently large yields $P(k)\sim k^{-\alpha}$, and the conditional Poisson (CP) distribution \cite{Clauset2009}
\begin{equation}\label{conditional-poisson}
\displaystyle P(k) = \left[ e^\mu -\sum_{m=0}^{k_\text{min}-1}\frac{\mu^m}{m!}\right]^{-1} \frac{\mu^k}{k!}, \quad k \geq k_{\text{min}},\ \mu > 0.
\end{equation}
Finally, we also considered a pairwise power law (PPL) distribution, which consists of two power law regions that are connected at $k=k_\text{trans}$:
\begin{equation}\label{pairwise-powerlaw}
\displaystyle P(k) = \left\lbrace \begin{split} &C \ k^{-\alpha}, & k_\text{min}\le k < \lceil k_\text{trans} \rceil \\       &C k_\text{trans}^{\beta-\alpha} \ k^{-\beta}, & \lceil k_\text{trans} \rceil  \le k
\end{split}\right., \quad \alpha,\beta>1,\ k_\text{trans}>k_\text{min},
\end{equation}
with the normalization factor $\displaystyle C= \left( \zeta(\alpha, k_\text{min})-\zeta(\alpha, \lceil k_\text{trans} \rceil) + 
k_\text{trans}^{\beta-\alpha} \zeta(\beta, \lceil k_\text{trans} \rceil) \right)^{-1}$. 
Due to the ceiling function $\lceil x \rceil$ this distribution does not strictly sum up to $1$. 
Still, as we will see in the following, it does provide in many instances a good fit to our data.

Inspection of these distributions reveal the presence of a minimal value $k_\text{min}$ that determines the start of the `tail' 
used for the modeling. In many cases $k_\text{min}$ can be determined as the value that minimizes the Kolmogorov-Smirnov statistics
between the empirical distributions and the fitted distributions \cite{Clauset2009}.

Due to the large size of our data we directly use the formula $\text{AIC} = - 2 \ln \hat{{\cal L}} + 2 n_p$ 
for the Akaike information criterion. Here $n_p$ is number of parameters in each distribution model 
and $\hat{{\cal L}}$ is the maximum likelihood of the model (see Appendix B).
The Akaike weight $w_i$ \cite{Burnham2002} for each model is
\begin{equation} \label{akaike-weight}
w_i = \frac{\exp\left((\text{AIC}_\text{min} - \text{AIC}_i)/2\right)}{\sum\limits_{j} 
\exp\left((\text{AIC}_\text{min} - \text{AIC}_j)/2\right)} ~,
\end{equation}
with the model with the largest Akaike weight being the most likely model.

Although we show in the following plots of the probability distribution functions (with logarithmic
binning) and of the complementary cumulative distribution functions as illustration, we do not 
use them directly for parameter estimation. Instead we estimate parameters from the distribution models
with the maximum likelihood method. The maximum likelihood estimators (MLE) for the different parameters
are summarized in Appendix B.

%

\section{Results}

\subsection{Search engine efficiency}

Before delving into a detailed analysis of foraging related quantities like step-lengths and waiting times,
we will first briefly characterize in a straightforward way the efficiency of the different search engines 
(or of the same search engine in different years) through 
the clicking number, i.e. the number of clicks needed before the user ends the query.

\begin{table}[!htbp] \label{table1}
\centering
\begin{tabular}{|c|ccc|}
\hline
        & Sogou-08 & Sogou-11 & Yahoo-10\\
\hline
year & $2008$ & $2011$ & $2010$\\
search engine provider & Sogou & Sogou & Yahoo\\
number of valid queries (in millions) & $14.6$ & $30.4$ & $53.8$ \\
$\langle N_c \rangle$ & $1.741$ & $1.433$ & $1.130$ \\
$\max(N_c)$ & $299$ & $23$ & $19$ \\
$P(N_c>10)$ & $0.626\%$ & $0.000033\%$ & $0.00655\%$\\
\hline
\end{tabular}
\caption{A comparison of the search engine efficiency based on the clicking number $N_c$. The third line
provides the total number of queries analyzed in our study. As a default the search engines provide 10 
links per page.}
\label{N_c_compare}
\end{table}

From the point of view of a user submitting a query to a search engine, what matters is the number of links they
have to click on before retrieving the needed information. The clicking number $N_c$ should therefore directly
reflect the efficiency of a search engine to provide the user with the relevant information.

Table I summarizes some of our findings for the clicking number. Inspection of the
table reveals immediately striking differences when comparing Sogou-08 with the other, newer, sets. Both the average
clicking number $\left< N_c \right>$ and the largest clicking number found in the millions of queries forming
the different sets point to an impressive increase of the efficiency when going from Sogou-08 to the newer sets.
The most striking difference is provided by the probability $P(N_c > 10)$ that more than 10 clicks are needed 
for accessing the relevant information. Whereas for Sogou-08 around 0.63\% of the queries result in more than 10 clicks
on links provided by the search engine, only a very small number of searches in Sogou-11 and Yahoo-10 result in the inspection
of more than 10 links.

\begin{figure} [h]
\includegraphics[width=0.70\columnwidth,clip=true]{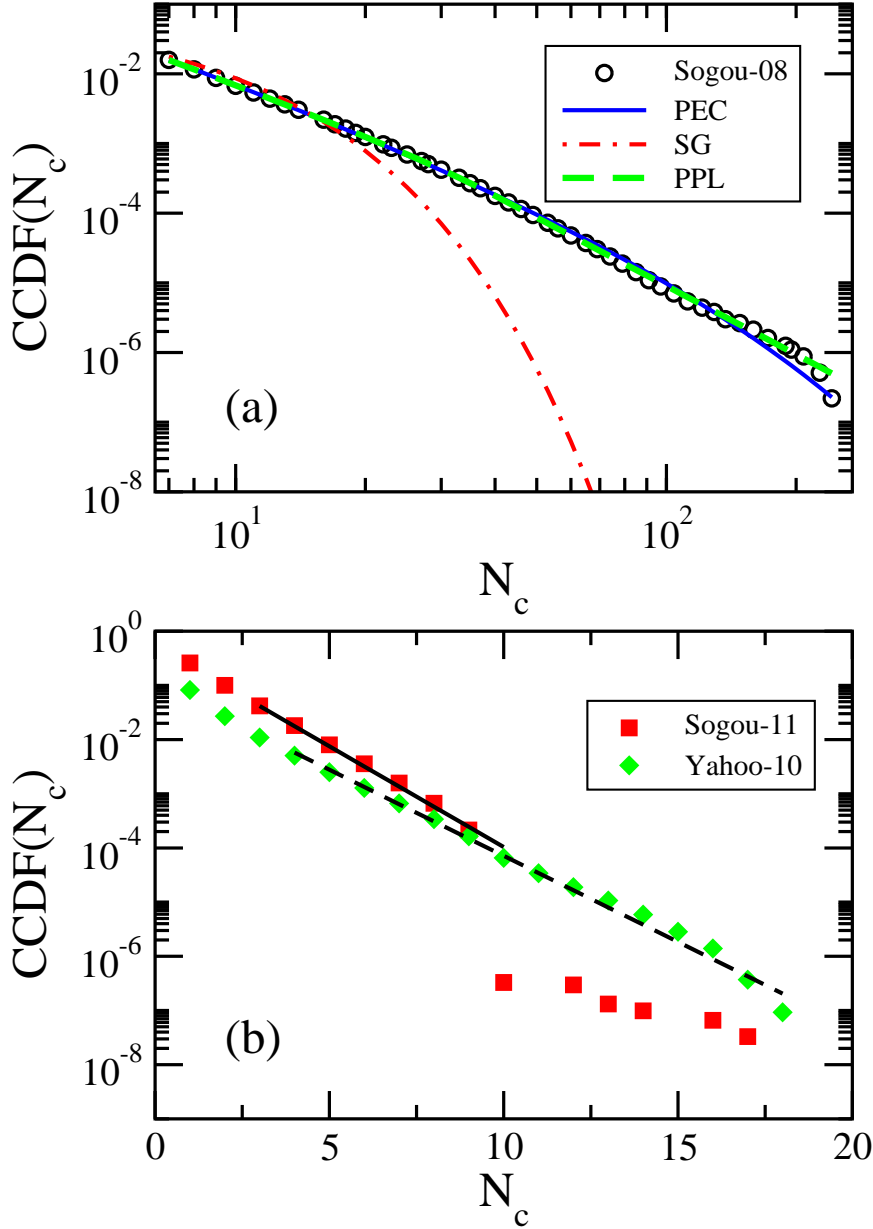}
\caption{\label{fig2}
(Color online) Complementary cumulative distribution function of the clicking number $N_c$ obtained from the different sets.
(a) Log-log plot of CCDF$(N_c)$ for Sogou-08 and comparison with the distributions obtained from three different models, see Table II:
power law with exponential cut-off (PEC), exponential model (SG), and pairwise power law (PPL). 
(b) Linear-log plot of CCDF$(N_c)$ for Sogou-11 and Yahoo-10. For Sogou-11
the probability distribution displays a dramatic drop for $N_c \ge 10$. The lines indicate an exponential decay.
}
\end{figure}

\begin{table*}[]
\centering
\begin{tabular}{|c|c|c|c|c|c|c|c|c|}
\hline
$\quad\ \quad$          & Set                       & model & $\ln\hat{{\cal L}}$ & AIC        & $w_i$   & most likely model                                                             & $k_\text{min}$          & MLE                                                                                                                            \\ \hline
\multirow{15}{*}{$N_c$} & \multirow{7}{*}{Sogou-08} & YS    & $-722852$           & $1445707$  & $0.000$ & \multirow{7}{*}{\begin{tabular}[c]{@{}c@{}}PPL\end{tabular}} & \multirow{7}{*}{7}  & \multirow{7}{*}{\begin{tabular}[c]{@{}c@{}}$\hat{\alpha}=3.488$\\ $\hat{\beta}=4.280$\\ $\hat{k}_\text{trans}=39.234$\end{tabular}}  \\ \cline{3-6}
                        &                           & DPL   & $-722912$           & $1445825$  & $0.000$ &                                                                               &                         &                                                                                                                                \\ \cline{3-6}
                        &                           & SG    & $-746099$           & $1492200$  & $0.000$ &                                                                               &                         &                                                                                                                                \\ \cline{3-6}
                        &                           & CP    & $-1000863$          & $2001728$  & $0.000$ &                                                                               &                         &                                                                                                                                \\ \cline{3-6}
                        &                           & DLN   & $-723126$           & $1446255$  & $0.000$ &                                                                               &                         &                                                                                                                                \\ \cline{3-6}
                        &                           & PEC   & $-722760$           & $1445524$  & $0.000$ &                                                                               &                         &                                                                                                                                \\ \cline{3-6}
                        &                           & PPL   & $-722739$           & $1445484$  & $1.000$ &                                                                               &                         &                                                                                                                                \\ \cline{2-9} 
                        & \multirow{4}{*}{Sogou-11} & YS    & $-3637855$          & $7275713$  & $0.000$ & \multirow{4}{*}{SG}                                                  & \multirow{4}{*}{$3$}    & \multirow{4}{*}{$\hat{\lambda}=0.855$}                                                                                         \\ \cline{3-6}
                        &                           & DPL   & $-3671768$          & $7343538$  & $0.000$ &                                                                               &                         &                                                                                                                                \\ \cline{3-6}
                        &                           & SG    & $-3603146$          & $7206293$  & $1.000$ &                                                                               &                         &                                                                                                                                \\ \cline{3-6}
                        &                           & CP    & $-3659310$          & $7318621$  & $0.000$ &                                                                               &                         &                                                                                                                                \\ \cline{2-9} 
                        & \multirow{4}{*}{Yahoo-10} & YS    & $-789652$           & $1579305$  & $0.000$ & \multirow{4}{*}{SG}                                                  & \multirow{4}{*}{$4$}    & \multirow{4}{*}{$\hat{\lambda}=0.732$}                                                                                         \\ \cline{3-6}
                        &                           & DPL   & $-794101$           & $1588203$  & $0.000$ &                                                                               &                         &                                                                                                                                \\ \cline{3-6}
                        &                           & SG    & $-786042$           & $1572086$  & $1.000$ &                                                                               &                         &                                                                                                                                \\ \cline{3-6}
                        &                           & CP    & $-803599$           & $1607199$  & $0.000$ &                                                                               &                         &                                                                                                                                \\ \hline
\multirow{7}{*}{$r_f$}  & \multirow{7}{*}{Sogou-08} & YS    & $-1657346$          & $3314695$  & $0.000$ & \multirow{7}{*}{\begin{tabular}[c]{@{}c@{}}PPL\end{tabular}} & \multirow{7}{*}{16} & \multirow{7}{*}{\begin{tabular}[c]{@{}c@{}}$\hat{\alpha}=2.108$\\ $\hat{\beta}=2.948$\\ $\hat{k}_\text{trans}=139.580$\end{tabular}} \\ \cline{3-6}
                        &                           & DPL   & $-1657674$          & $3315350$  & $0.000$ &                                                                               &                         &                                                                                                                                \\ \cline{3-6}
                        &                           & SG    & $-1734486$          & $3468975$  & $0.000$ &                                                                               &                         &                                                                                                                                \\ \cline{3-6}
                        &                           & CP    & $-9745388$          & $19490777$ & $0.000$ &                                                                               &                         &                                                                                                                                \\ \cline{3-6}
                        &                           & DPL   & $-1655263$          & $3310530$  & $0.000$ &                                                                               &                         &                                                                                                                                \\ \cline{3-6}
                        &                           & PEC   & $-1654553$          & $3309109$  & $0.000$ &                                                                               &                         &                                                                                                                                \\ \cline{3-6}
                        &                           & PPL   & $-1654359$          & $3308723$  & $1.000$ &                                                                               &                         &                                                                                                                                \\ \hline
\end{tabular}\label{mle-aic-Nc-rf}
\caption{Model selection using AIC and maximum likelihood estimators for the parameters in the most likely models of $N_c$ and $r_f$.
In the table $\ln\hat{\mathcal{L}}$ and AIC are rounded to integers. See the main text for the meaning of the acronyms.}
\end{table*}

\begin{table*}[]
\centering
\begin{tabular}{|c|c|c|c|c|c|c|c|c|}
\hline
$\quad\ \quad$                  & Set                       & model & $\ln\hat{{\cal L}}$ & AIC        & $w_i$   & most likely model                                                             & $k_\text{min}$          & MLE                                                                                                                          \\ \hline
\multirow{15}{*}{$d$}           & \multirow{7}{*}{Sogou-08} & YS    & $-2090805$          & $4181612$  & $0.000$ & \multirow{7}{*}{\begin{tabular}[c]{@{}c@{}}PPL\end{tabular}} & \multirow{7}{*}{10} & \multirow{7}{*}{\begin{tabular}[c]{@{}c@{}}$\hat{\alpha}=2.169$\\ $\hat{\beta}=3.417$\\ $\hat{k}_\text{trans}=91$\end{tabular}}    \\ \cline{3-6}
                                &                           & DPL   & $-2091568$          & $4183137$  & $0.000$ &                                                                               &                         &                                                                                                                              \\ \cline{3-6}
                                &                           & SG    & $-2187409$          & $4374821$  & $0.000$ &                                                                               &                         &                                                                                                                              \\ \cline{3-6}
                                &                           & CP    & $-7645269$          & $15290541$ & $0.000$ &                                                                               &                         &                                                                                                                              \\ \cline{3-6}
                                &                           & DLN   & $-2088146$          & $4176295$  & $0.000$ &                                                                               &                         &                                                                                                                              \\ \cline{3-6}
                                &                           & PEC   & $-2087035$          & $4174075$  & $0.000$ &                                                                               &                         &                                                                                                                              \\ \cline{3-6}
                                &                           & PPL   & $-2085243$          & $4170491$  & $1.000$ &                                                                               &                         &                                                                                                                              \\ \cline{2-9} 
                                & \multirow{4}{*}{Sogou-11} & YS    & $-22684073$         & $45368149$ & $0.000$ & \multirow{4}{*}{SG}                                                  & \multirow{4}{*}{$1$}    & \multirow{4}{*}{$\lambda=0.544$}                                                                                             \\ \cline{3-6}
                                &                           & DPL   & $-23284589$         & $46569181$ & $0.000$ &                                                                               &                         &                                                                                                                              \\ \cline{3-6}
                                &                           & SG    & $-21326460$         & $42652922$ & $1.000$ &                                                                               &                         &                                                                                                                              \\ \cline{3-6}
                                &                           & CP    & $-22894455$         & $45788912$ & $0.000$ &                                                                               &                         &                                                                                                                              \\ \cline{2-9} 
                                & \multirow{4}{*}{Yahoo-10} & YS    & $-12128059$         & $24256120$ & $0.000$ & \multirow{4}{*}{SG}                                                  & \multirow{4}{*}{$1$}    & \multirow{4}{*}{$\lambda=0.540$}                                                                                             \\ \cline{3-6}
                                &                           & DPL   & $-12450282$         & $24900567$ & $0.000$ &                                                                               &                         &                                                                                                                              \\ \cline{3-6}
                                &                           & SG    & $-11415241$         & $22830484$ & $1.000$ &                                                                               &                         &                                                                                                                              \\ \cline{3-6}
                                &                           & CP    & $-12326037$         & $24652076$ & $0.000$ &                                                                               &                         &                                                                                                                              \\ \hline
\multirow{4}{*}{$d_\text{in}$}  & \multirow{4}{*}{Sogou-08} & YS    & $-16506000$         & $33012001$ & $0.000$ & \multirow{4}{*}{SG}                                                  & \multirow{4}{*}{$1$}    & \multirow{4}{*}{$\lambda=0.550$}                                                                                             \\ \cline{3-6}
                                &                           & DPL   & $-16924868$         & $33849738$ & $0.000$ &                                                                               &                         &                                                                                                                              \\ \cline{3-6}
                                &                           & SG    & $-15652590$         & $31305183$ & $1.000$ &                                                                               &                         &                                                                                                                              \\ \cline{3-6}
                                &                           & CP    & $-16985431$         & $33970864$ & $0.000$ &                                                                               &                         &                                                                                                                              \\ \hline
\multirow{7}{*}{$d_\text{out}$} & \multirow{7}{*}{Sogou-08} & YS    & $-1232283$          & $2464569$  & $0.000$ & \multirow{7}{*}{\begin{tabular}[c]{@{}c@{}}PPL\end{tabular}} & \multirow{7}{*}{1}  & \multirow{7}{*}{\begin{tabular}[c]{@{}c@{}}$\hat{\alpha}=2.353$\\ $\hat{\beta}=3.226$\\ $\hat{k}_\text{trans}=9.000$\end{tabular}} \\ \cline{3-6}
                                &                           & DPL   & $-1225123$          & $2450249$  & $0.000$ &                                                                               &                         &                                                                                                                              \\ \cline{3-6}
                                &                           & SG    & $-1435452$          & $2870906$  & $0.000$ &                                                                               &                         &                                                                                                                              \\ \cline{3-6}
                                &                           & CP    & $-1943098$          & $3886197$  & $0.000$ &                                                                               &                         &                                                                                                                              \\ \cline{3-6}
                                &                           & DLN   & $-1235756$          & $2471516$  & $0.000$ &                                                                               &                         &                                                                                                                              \\ \cline{3-6}
                                &                           & PEC   & $-1223417$          & $2446838$  & $0.000$ &                                                                               &                         &                                                                                                                              \\ \cline{3-6}
                                &                           & PPL   & $-1221825$          & $2443655$  & $1.000$ &                                                                               &                         &                                                                                                                              \\ \hline
\end{tabular}\label{mle-aic-d}
\caption{Model selection using AIC and maximum likelihood estimators for the parameters in the most likely model for the step-lengths.
See the main text for the meaning of the acronyms.}
\end{table*}

The distribution CCDF($N_c$) shown in Fig. \ref{fig2} reveals the dramatic differences between the efficiency
(as measured by $N_c$) of the different search engines. As reported in Table II, we find using AIC that 
the probability distribution for Sogou-08
follows a pairwise power law distribution with the maximum likelihood 
estimations $\hat{\alpha}=3.448, \hat{\beta}=4.280$, whereas the transition happens at $\hat{k}_\text{trans}=39.234$.
For Sogou-11 and Yahoo-10, however, the distributions rapidly show an exponential decay. This exponential
decay stops at $N_c=10$ for Sogou-11, indicating that with a very few exceptions all queries are finished 
within 10 clicks. 
This discontinuity does not happen in Yahoo-10 which instead shows a smooth behavior. The jump in Sogou-11 reveals that
with a few exceptions users found within the links provided on the very first page with results the information
they were looking for.

\begin{figure} [h]
\includegraphics[width=0.70\columnwidth,clip=true]{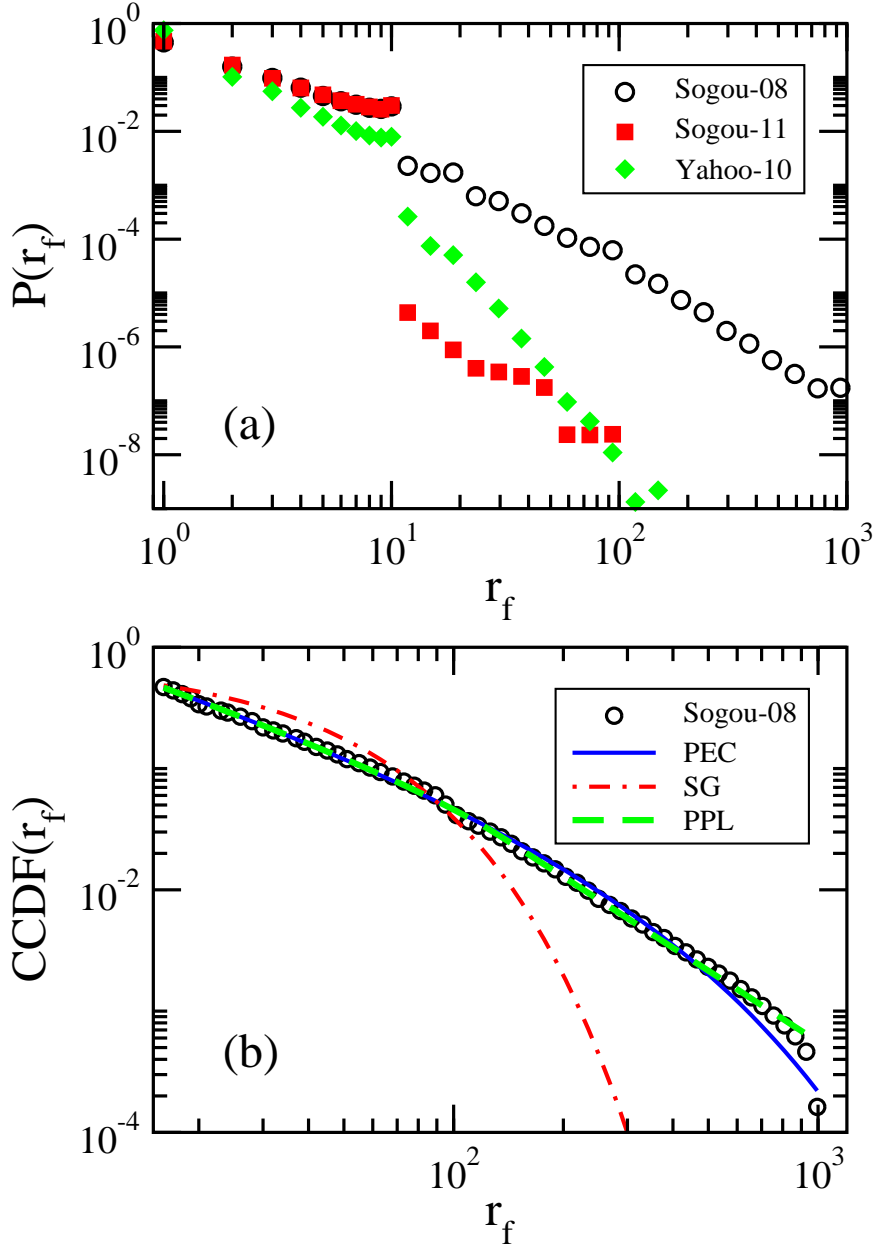}
\caption{\label{fig3}
(Color online) (a) Probability distribution $P(r_f)$ where $r_f$ is the rank of the link that is clicked
as last in a given search. $r_f$ therefore corresponds to the position of the resource on the semi-infinite line. 
Large differences can be observed between the tails of Sogou-08 and Sogou-11. (b) Complementary cumulative
distribution function CCDF$(r_f)$ for Sogou-08 as well as the corresponding distributions obtained from three different models, see Table II:
power law with exponential cut-off (PEC), exponential model (SG), and pairwise power law (PPL).
}
\end{figure}

One also expects from an efficient search engine that the relevant result is included in the very first
suggested links. Fig. \ref{fig3} compares for our three data sets the probability distribution $P(r_f)$ 
where $r_f$ is the rank of the final click (i.e. the position of the resource on the semi-infinite line).
We first note that for Yahoo-10 75\% of the searches end with a click on the very first link on the search
result page provided by the search engine. For Sogou-11 this number is 47\%, very similar to the 45\% of searches
that end with a click on the very first link for Sogou-08. For Sogou-11 and Yahoo-10 almost all resources (99.997\% for Sogou-11 and
99.864\% for Yahoo-10) are located on the first page with $1 \leq r_f \leq 10$. For these two cases one also
observes large changes between $P(r_f=10)$ and $P(r_f=11)$, illustrating the fact that only for a negligible
number of searches the resource is found for $r_f > 10$. This is different for Sogou-08 where $P(r_f > 10)=3.661\%$
and $P(r_f > 100) = 0.224\%$, resulting in a much smoother shape with a pairwise power-law tail.
The probability distribution $P(r_f)$ indicates
that for Sogou-11 and Yahoo-10 only a local exploration
on the first page is needed in order to reach the resource. As we will see in the following, this difference
yields different space-time patterns during the searches.

\subsection{Step-lengths and waiting times}

\begin{figure} [h]
\includegraphics[width=0.70\columnwidth,clip=true]{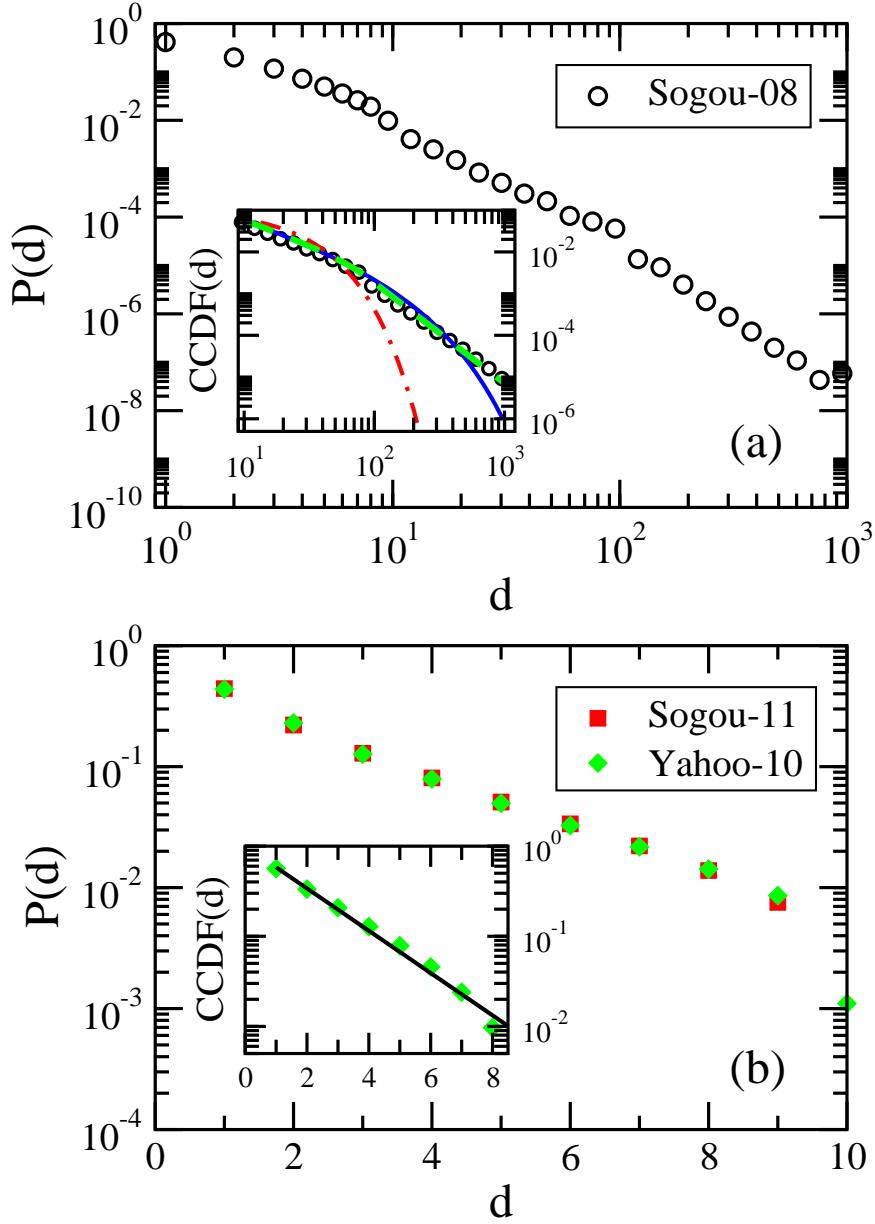}
\caption{\label{fig4}
(Color online) Step-length probability distribution functions and (in the inset) complementary cumulative distribution functions. 
(a) For Sogou-08 the distributions
display a pairwise power-law tail, whereas
(b) for Sogou-11 and Yahoo-10 they decay exponentially, see Table III. The lines in the inset of panel (a) show the
following fitted models: PEC (solid blue line), SG (dot-dashed red line), and PPL (dashed green line). The solid black line in
the inset of panel (b) represents an exponentially decaying function.}
\end{figure}

In the following analysis we view as a random walk on the semi-infinite line
the exploration by the user of the links provided by the search engine.
Focusing on the step-length and waiting
time distributions, we will see that the difference in efficiency noticed in the previous subsection
yields different space-time processes.

A way to distinguish between
Brownian-type motion and L\'{e}vy movement 
is to investigate the probability distribution $P(d)$ of the step-length $d$, 
which should display a heavy tail in the form of a power-law
\begin{equation} \label{eq:Levy}
P(d) \sim d^{-\alpha}
\end{equation}
with $1 < \alpha < 3$ for 
super-diffusive L\'{e}vy flights.

\begin{figure} [h]
\includegraphics[width=0.70\columnwidth,clip=true]{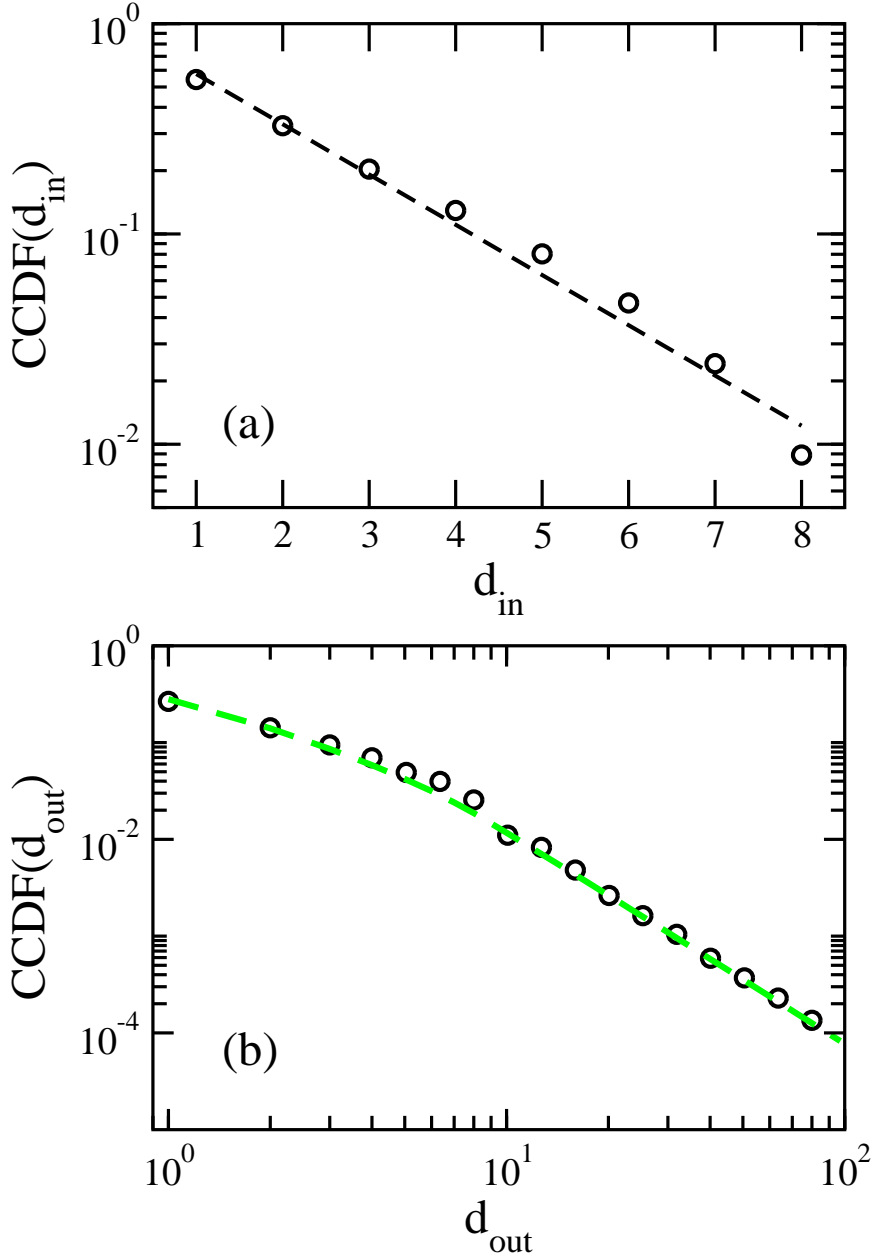}
\caption{\label{fig5}
(Color online) (a) Step-length distribution for jumps within a page for Sogou-08. An exponential decay is observed 
for local searches, similarly to the distributions encountered for Sogou-11 and Yahoo-10. (b)
Distribution of the page difference
for jumps between pages for Sogou-08. From the model selection follows that these data are described
by a pairwise power law distribution, indicated by the green dashed line (see also Table III).}
\end{figure}

Fig. \ref{fig4} shows the probability distributions for the step-length derived from the data at our disposal.
Focusing first on Sogou-08, our model selection procedure shows that the pairwise power law model provides a good absolute fit,
see Fig. \ref{fig4}a as well as Fig. \ref{fig5}b. From the maximum likelihood estimation we obtain that the transition between the two power laws happens
at $d_\text{trans}=91$. The exponent in the first power-law region is given by $\alpha=2.169$, a value that is 
within the L\'{e}vy-flight range $1<\alpha<3$, i.e. between $10 < d < d_\text{trans}$ we have a long-range search pattern that
is consistent with the power law distribution of a L\'{e}vy flight.
This behavior does not persists for the largest values of $d$. Instead, the exponent in the second power-law region is $\beta=3.417$,
which is outside the L\'{e}vy-flight range. This value guarantees a finite variance for step-lengths and suggests that 
the very long-range movements have the properties of normal diffusion. We believe that this change in behavior 
around $d_\text{trans}=91$ is due to the layout of the search engine result pages, since the search engines used in our study 
list $10$ pages at the bottom of a result page.
For Sogou-11 and Yahoo-10, 99.997\% and 99.890\% of steps have a length $d < 10$. The distributions for $d < 10$ are
exponential which 
points to the overwhelming predominance of local searches where only a small ``area" is
explored. 

Interestingly, even for Sogou-08 an exponential decay is hidden in the distribution shown in Fig. \ref{fig4}a. 
Separating the jumps within a page from those between pages,
we discover in Fig.~\ref{fig5} a more complex behavior. Restricting ourselves to jumps within a given page,
where we denote by $d_{in}$ the corresponding step-length, we find for Sogou-08 an exponential decay
of the step-length distribution. The difference between Sogou-08 and the other sets therefore
mainly results from searches where the resource is not readily found, yielding jumps between different
pages with a pairwise power-law probability distribution for the out-of-page step-length $d \ge 10$. 
As shown in Fig.~\ref{fig5}b, the page difference $d_{out}$ of out-of-page jumps 
still yields a pairwise power law distribution.


The fact that Sogou-08 yields a switch between a local (i.e. on one page with search results) Brownian search and 
a relocation phase that is power law distributed is very reminiscent of an intermittent search process that includes
L\'{e}vy strategies \cite{Lon08,Ben11}. Intermittent search processes have been proposed as search strategies in
cases where the targets are hidden \cite{Ben05,Ben06,Osh09}. They are characterized by switches between two different phases:
careful searches around one location, followed by rapid relocations to some
other areas. The careful searches are usually described as Brownian searches whereas the relocations are often
assumed to be either ballistic or L\'{e}vy distributed. The set-up of search engines queries has many obvious
direct connections with an intermittent search process. This is especially true for Sogou-08 for which we observe
relocations over large distances. 

\begin{figure} [h]
\includegraphics[width=0.50\columnwidth,clip=true]{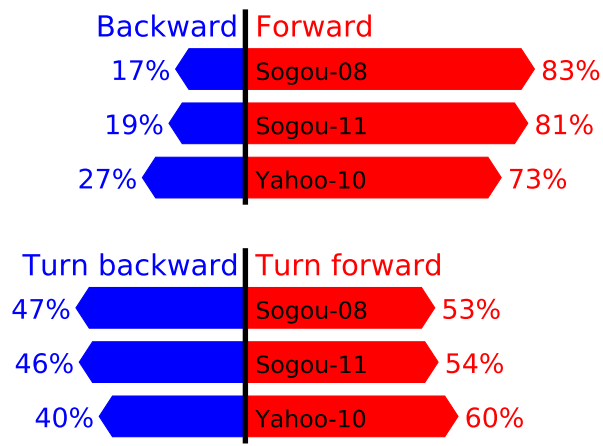}
\caption{\label{fig6}
(Color online) Upper part: fraction of jumps in the forward and backward directions. Lower part: 
fraction of turns that change the direction from forward to backward and of turns that change the
direction from backward to forward.
}
\end{figure}

For any of the processes usually discussed in the context of foraging, one generally assumes the movement to
be unbiased, i.e. that jumps are happening with a direction independent probability distribution. As we know from
our own experience and as shown in Fig. \ref{fig6} for the different data sets, this is not the case in online searches,
where users have a tendency to start at the top of a page with research results and proceed 
to the bottom of the page (i.e. to move preferentially in one direction, see top of Fig. \ref{fig6}). 
While there is a clear directionality in how a user exploits search results,
a much smaller bias is observed in the 'turning angles', i.e. in the changes of direction from forward to backward and
from backward to forward.

On the semi-infinite line, which provides the landscape for online foraging,
moving forward usually means exploring new results, while moving backwards often
means revisiting previously viewed results. The bias of foraging means that there
is much more exploration of new results than revisitation of already viewed ones.
Meanwhile, since most of the initial movements are forward when users start to
view results from a search, the much weaker bias in turning angles indicates that
a revisitation is usually followed by another exploration.

\begin{figure} [h]
\includegraphics[width=0.70\columnwidth,clip=true]{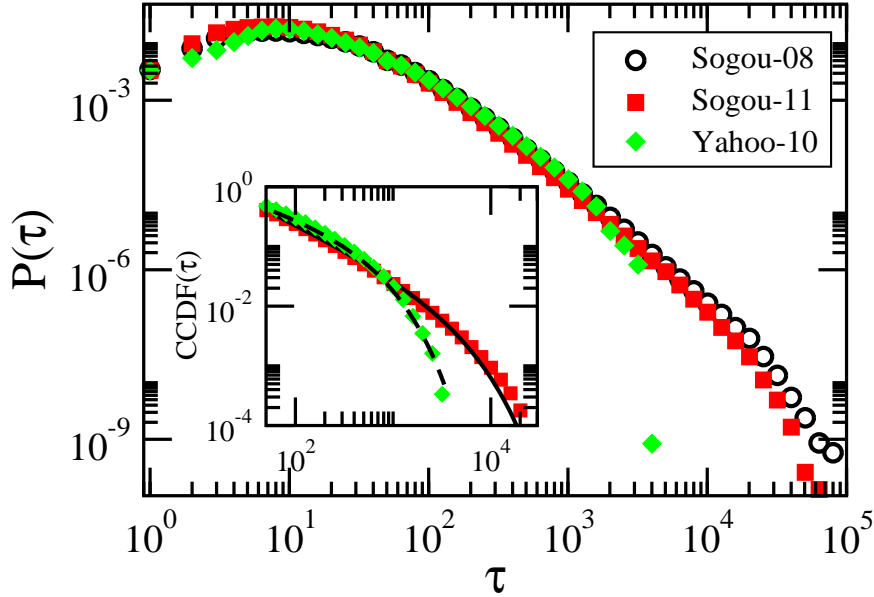}
\caption{\label{fig7}
(Color online) Waiting time distributions $P(\tau)$ for the different sets,
with $\tau$ measured in seconds. For all three cases the distribution exhibits a
power-law tail, with an exponent around 1.9, followed by an exponential cut-off. The different sets 
have different upper limits for the waiting times: one day for Sogou-08 and Sogou-11 and one hour for Yahoo-10.
Inset: the corresponding complementary cumulative distribution functions and the fits (solid and dashed lines) obtained from 
power law models with exponential cut-offs.
}
\end{figure}

Other differences between online searches and models for foraging emerge when considering the 
probability distribution $P(\tau)$ of the waiting time $\tau$, i.e.
the time elapsed between two consecutive clicks on links provided by the search engine. 
The time between two clicks on the links provided by the search engine is mainly the time spent by
the user viewing a web site. The average time spent on selecting the next link on the search result pages
is indeed small compared to the average time spent on a selected web site.
Inspection of Fig. \ref{fig7} reveals that $P(\tau)$ is well modeled by a power law distribution with exponential cut-off, see inset.
This is especially true for Yahoo-10 where our model selection indeed finds that a power law distribution with exponential cut-off
provides the best fit. For Sogou-08 and Sogou-11 AIC yields the log-normal distribution as the most probable one, but the
log-normal distribution does not at all capture the behavior for large $\tau$, which instead is well described by a
power law distribution with exponential cut-off, see the black lines in the inset of Fig. \ref{fig7}. This exponential
cut-off in $P(\tau)$ is due to upper limits for the waiting time set by the session expiration time (one hour for Yahoo-10
and one day for Sogou-08 and Sogou-11).

\subsection{Mean square displacement}

For Brownian motion and L\'{e}vy flights the mean square displacement shows a characteristic behavior,
increasing linearly with the number of jumps for the first case, whereas for the second case a super-diffusive
behavior is expected, with a power-law increase where the exponent is larger than one. 

\begin{figure} [h]
\includegraphics[width=0.70\columnwidth,clip=true]{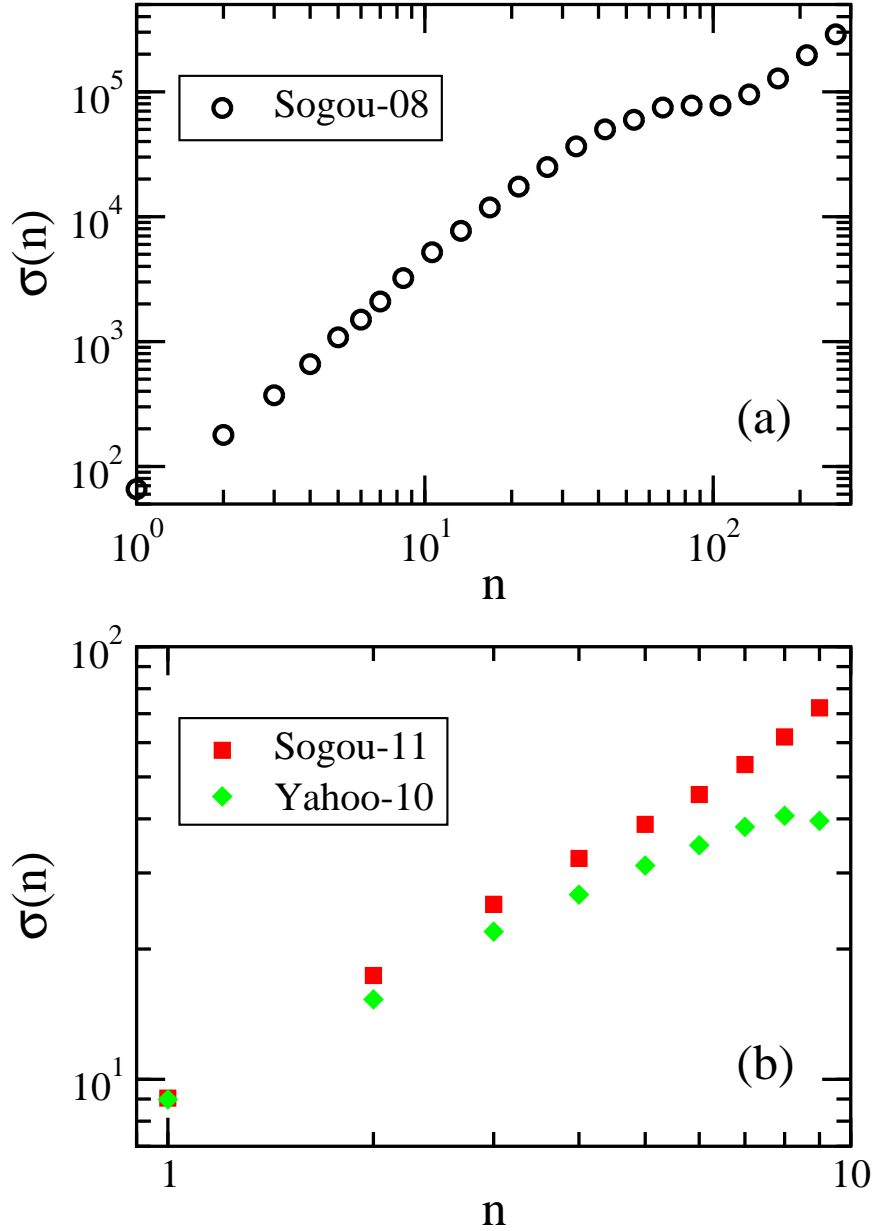}
\caption{\label{fig8}
(Color online) Mean square displacement $\sigma(n)$ as a function of the clicking order $n$. (a) A super-diffusive
behavior is observed for Sogou-08. (b) For Sogou-11 and Yahoo-10 a slightly sub-diffusive
behavior is revealed by $\sigma(n)$.
}
\end{figure}

Regarding a query again as a motion along the semi-infinite line where the rank of a click $r$ corresponds
to the position whereas the clicking order $n$ counts the
number of jumps and therefore serves as a proxy for ``time", we can calculate the mean square displacement
as
\begin{equation} \label{sigma_n}
\sigma(n) = \langle \left( r(n) - r(1) \right)^2 \rangle
\end{equation}
where $r(1)$ is the rank of the first clicked result. In (\ref{sigma_n}) the average is over the different trajectories along
the semi-infinite line.

Fig. \ref{fig8} shows for the different sets the variation of the mean square displacement with increasing
clicking order $n$. For Sogou-08, see Fig. \ref{fig8}a, we do find for $n \leq 30$ the expected super-diffusive behavior,
$\sigma(n) \sim n^a$, with an exponent $a \approx 1.95$. For Sogou-11 and Yahoo-10 reliable data are only available for
small values of $n$ due to the fact that the resource is usually found after only a few clicks. For $n < 10$
we find for Sogou-11 an exponent $a=0.92$, whereas for Yahoo-10 the value of the exponent is $a = 0.75$.
These values, which indicate a slight sub-diffusive behavior, are rather close to the value $a=1$ of normal
diffusion. 

\begin{figure} [h]
\includegraphics[width=0.70\columnwidth,clip=true]{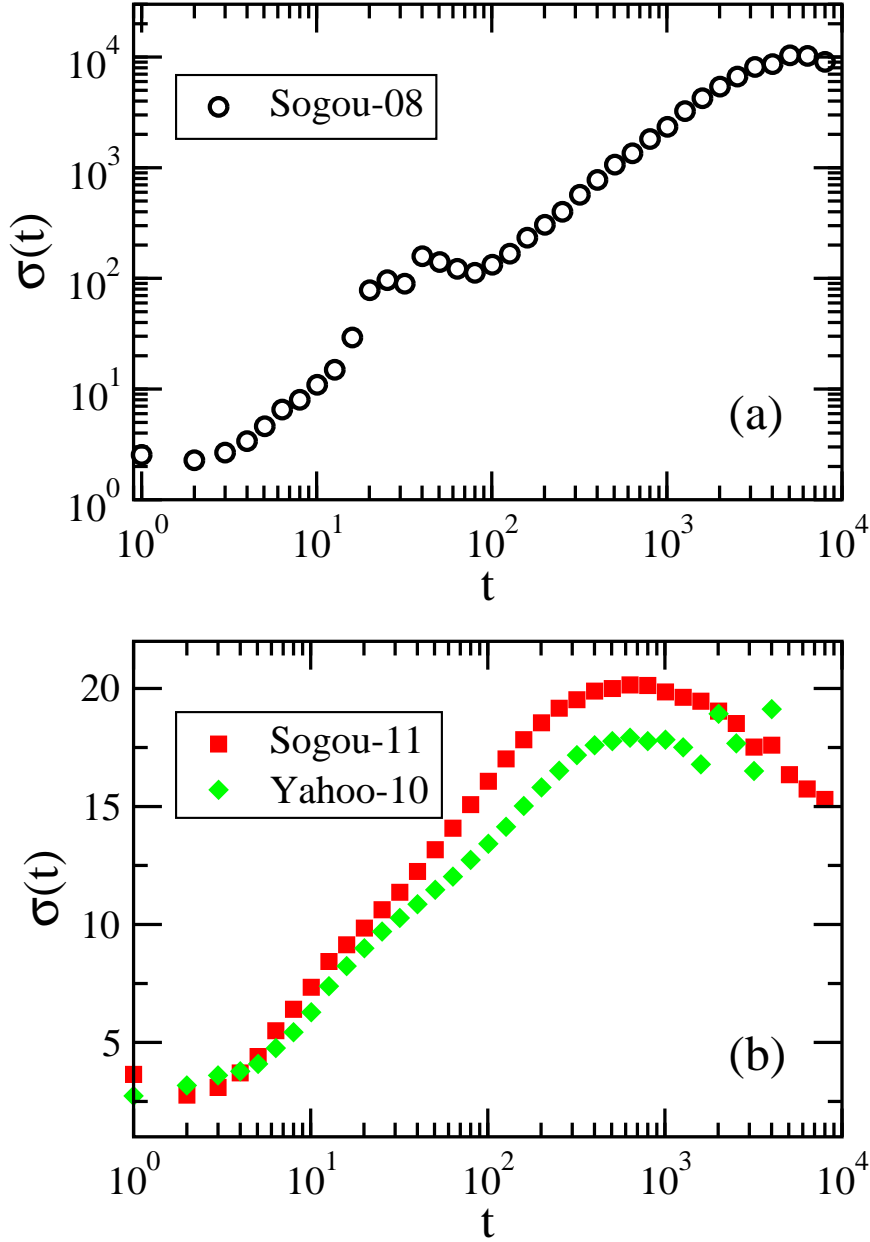}
\caption{\label{fig9}
(Color online) Mean square displacement $\sigma(t)$ as a function of time $t$ (measured in seconds) 
elapsed since the very first click on
a link provided by the search engine. Whereas for Sogou-08 a power law with an exponent close to 1.30 is observed in the
time interval 100 $s$ $< t <$ 2000 $s$, for Sogou-11 and Yahoo-10 $\sigma(t)$ is found to vary logarithmically with time 
for 10 $s$ $< t <$ 500 $s$.
}
\end{figure}

As we know from the click-through logs the time elapsed between any two consecutive clicks,
we can also calculate the mean square displacement as a function of the real time measured since
the very first click:
\begin{equation} \label{sigma_t}
\sigma(t) = \langle \left( r(t) - r_0) \right)^2 \rangle
\end{equation}
where $r_0$ is the rank of the first click at time $t=0$. We know from Fig. \ref{fig7} that
for all three sets the distributions of waiting times, which are composed by the times spent on selecting the
next link on the search result pages and the times spent viewing the previously selected website,
are rather complicated. 
This will of course impact the time dependence of $\sigma(t)$. Our result for Sogou-08 shown in Fig. \ref{fig9}a 
indicates that for that case the time-dependent mean square displacement varies in the time
interval 100 $s$ $< t <$ 2000 $s$ like a power law with an exponent close to 1.30. 
For Sogou-11 and Yahoo-10, however, $\sigma(t)$ roughly varies logarithmically 
with time in the interval 10 $s$ $< t <$ 500 $s$, see Fig. \ref{fig9}b. 

\subsection{Entropy}

\begin{figure} [h]
\includegraphics[width=0.30\columnwidth,clip=true]{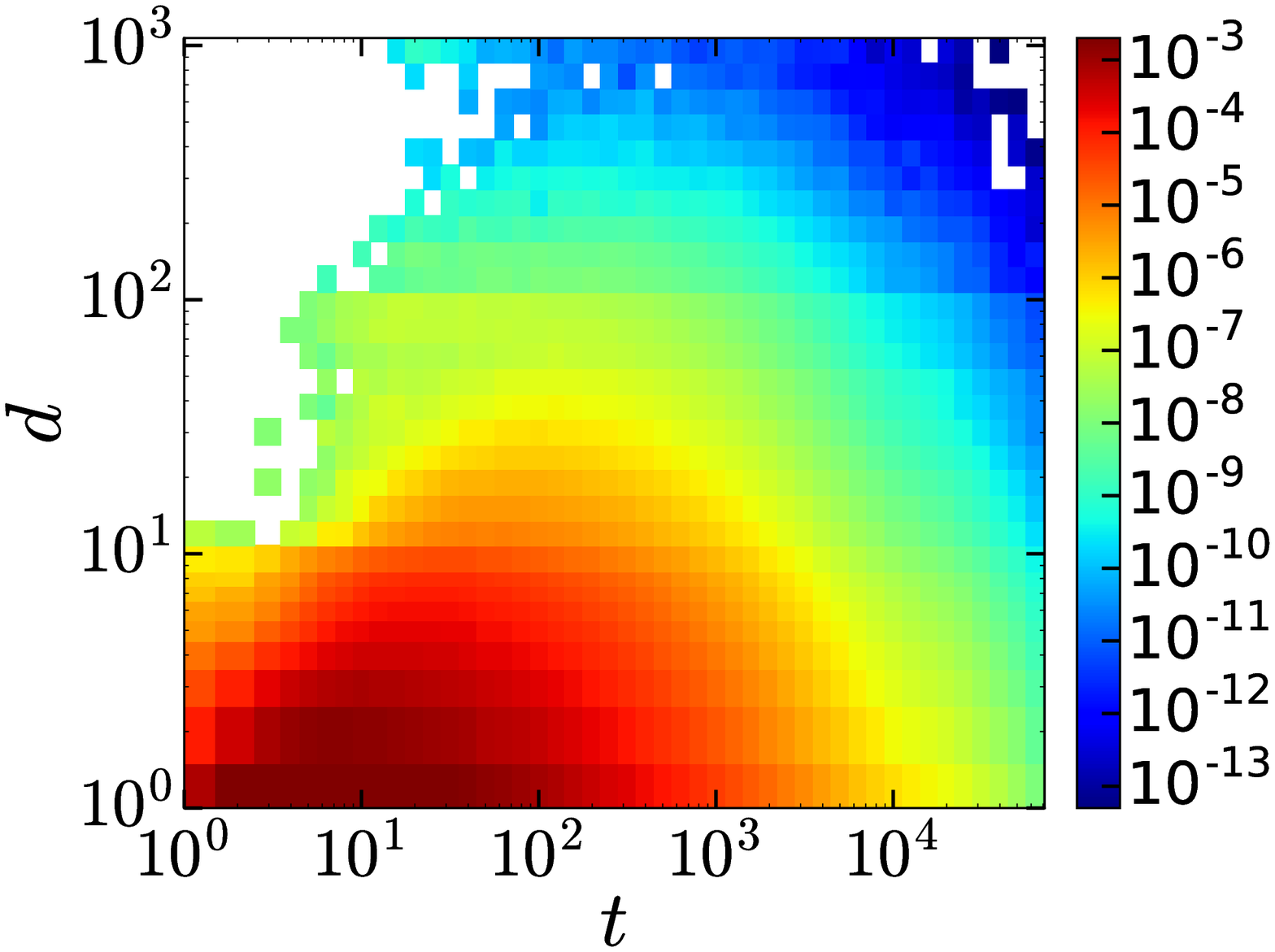}~~
\includegraphics[width=0.30\columnwidth,clip=true]{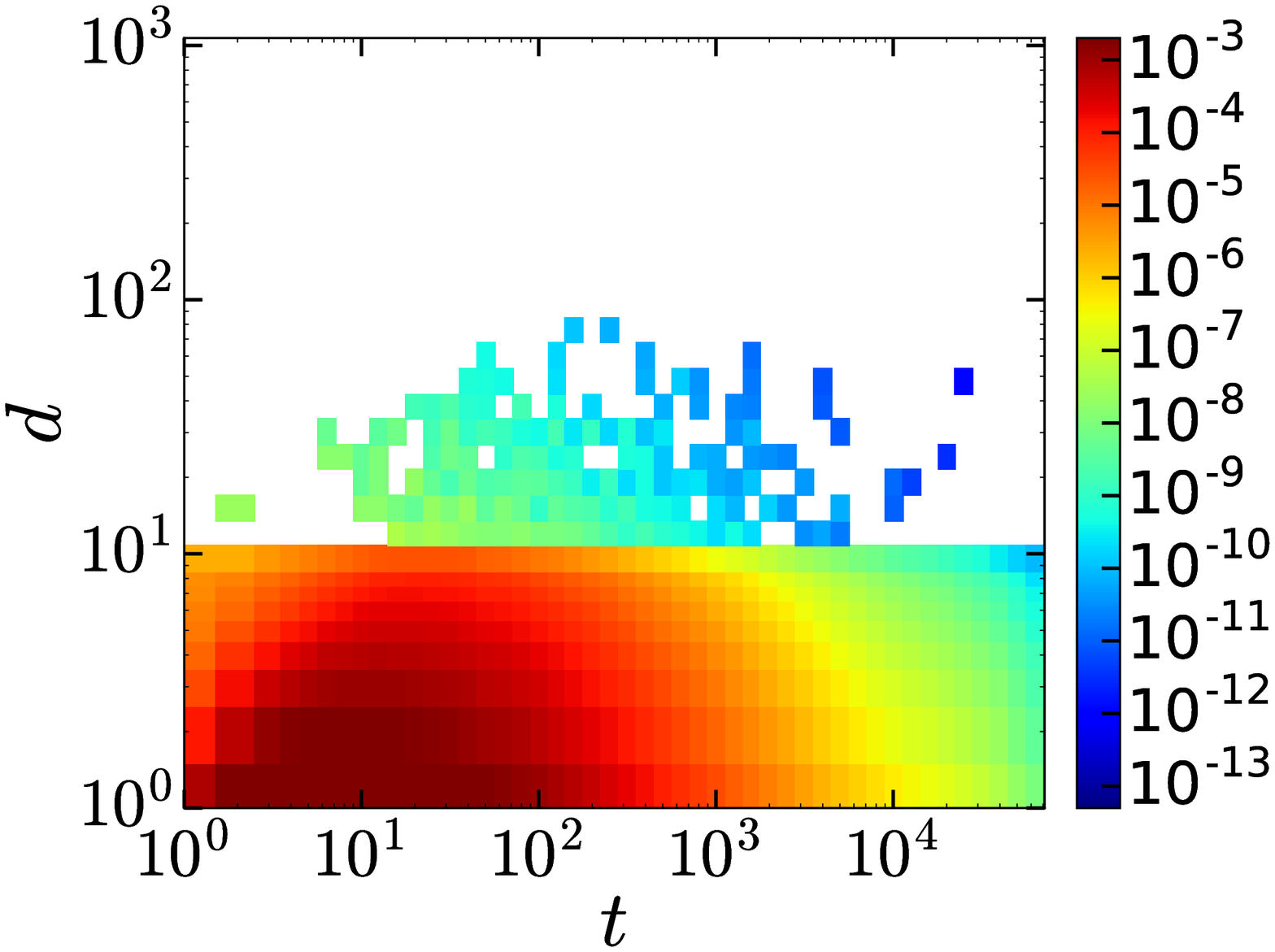}~~
\includegraphics[width=0.30\columnwidth,clip=true]{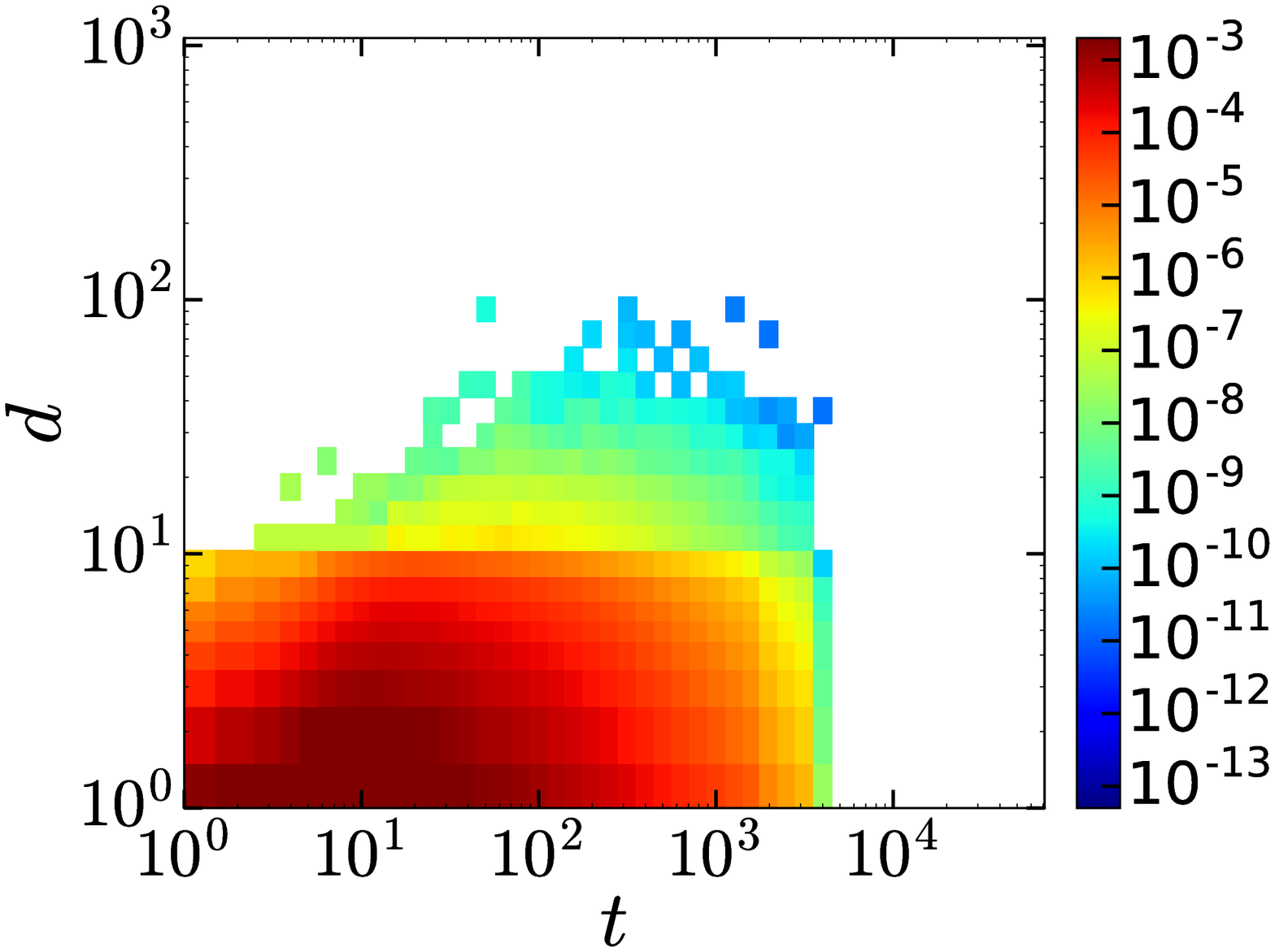}
\caption{\label{fig10}
(Color online) Joint probability $P(t,d)$ for Sogou-08 (left panel), Sogou-11 (middle panel),
and Yahoo-10 (right panel).
Note the sharp transition at $d=10$ for Sogou-11 and Yahoo-10, due to the rather few jumps between different pages.
$t$ is the time measured in seconds since the first click on a link and $d$ is the step-length between
two consecutive clicks.
}
\end{figure}

The differences between the different sets can also be highlighted through the study of time-dependent
entropies. We can for example start from the conditional probability $P(d|t)=P(t,d)/P(t)$, which is the probability that the step-length 
is $d$ given that the jump takes place at time $t$.
Here $P(t,d)$ is the joint probability of the time $t$ elapsed since the
first click and the step-length $d$, whereas $P(t)$ is the probability that the event happens at time $t$. 
This allows us to define the time-dependent entropy 
\begin{equation} \label{eq:ent}
S_d(t) = - \sum\limits_{d=1}^{\infty} P(d|t) \ln P(d|t)
\end{equation}
We could also start from a different conditional probability distribution, e.g. $P(d|n)$ where $n$ is the clicking order, 
\begin{equation} \label{eq:enn}
S_d(n) = - \sum\limits_{d=1}^{\infty} P(d|n) \ln P(d|n)~.
\end{equation}

Fig. \ref{fig10} compares the joint probabilities $P(t,d)$ for the different sets.
We note again that for Sogou-11 and Yahoo-10 almost all displacements are local with $d < 10$,
which is reflected by the very low or even vanishing joint probability $P(t,d)$ for $d \ge 10$ for these cases.
We also note that for $d \leq 9$ the probabilities are very similar for the different sets, which indicates 
that the properties of local searches are rather set independent. The probability 
for Sogou-08 reveals the emergence and distribution of long-range relocations.
Finally, for all cases $P(t,d)$ changes as a function of time, as expected for
a search process that is taking place far from equilibrium.

\begin{figure} [h]
\includegraphics[width=0.70\columnwidth,clip=true]{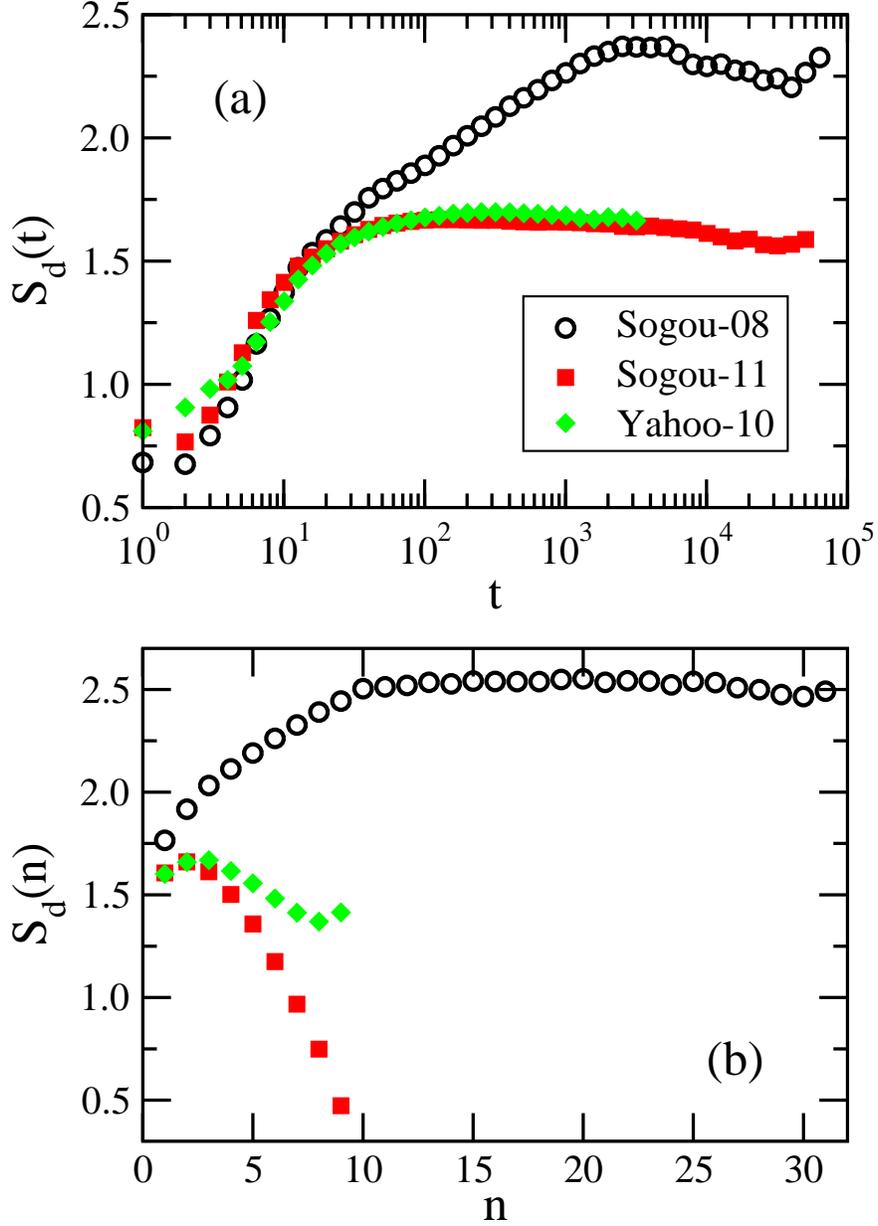}
\caption{\label{fig11}
(Color online) (a) Time dependence of the entropy $S_d(t)$ for the different sets. The early time regime is
characterized by a strong increase of $S_d(t)$. For Sogou-11 and Yahoo-10 a plateau is reached much earlier
than for Sogou-08 (of the order of a minute for Sogou-11 and Yahoo-10 and of the order of an hour for Sogou-08). 
For Sogou-08 the increase of the entropy at intermediate times is due to the predominance
of long relocation jumps in this regime. (b) Entropy $S_d(n)$ as a function of the clicking order $n$.
For Sogou-08, $S_d(n)$ reaches a plateau for $n \geq 11$.
}
\end{figure}

The time dependence of the entropy $S_d(t)$ defined in Eq. (\ref{eq:ent}) is shown in Fig. \ref{fig11}a.
We first note that for all three sets the entropy shows a strong increase at early times. For Sogou-11 and Yahoo-10,
characterized by mostly local searches on the first page with links, this increase rapidly weakens and 
$S_d$ reaches a plateau for times $t > 50$ seconds. For Sogou-08, where the motion is formed by 
a combination of local searches and long relocation jumps, the entropy keeps increasing up to $t \approx 2000$ seconds
before reaching a plateau. $S_d(t)$ is therefore another quantity that allows to easily distinguish between
Brownian-like searches and searches that are characterized by power law distributions.

Fig. \ref{fig11}b shows the entropy $S_d(n)$ defined in Eq. (\ref{eq:enn}). For Sogou-08, $S_d(n)$, after an initial increase
for small $n$, rapidly reaches a plateau for $n \geq 11$. The process is therefore stationary for $n \geq 11$. 
In contrast to this, for Sogou-11 and Yahoo-10 the entropy does not reach a well defined plateau, and the processes
therefore do not reach a stationary state.

\subsection{Correlations}
Correlation coefficients allow us to gain additional insights into the relationships between different 
quantities (see Appendix C for the definitions of the correlation coefficients discussed in the following). 
This is of interest as Brownian motions and L\'{e}vy flights, which are commonly used to model
human dynamics, assume that space and time are uncorrelated and that the random walks are memoryless. Studying
correlation coefficients will allow us to see to what extend these assumptions are fulfilled
in human online search processes.


Well suited correlation coefficients for our purpose are Kendall's tau \cite{Ken90}, $\tau_K$,
and Spearman's rho \cite{Tay87}, $\rho_S$, as these two non-parametric
measures are distribution-free and can handle power-law distributed quantities. Both coefficients are
rank-based and measure the correspondence of two series of ordinal numbers. Series for waiting times and
displacements are of course readily obtained from the original search engine click-through logs.

\begin{table}[!htbp]
\centering
\begin{tabular}{|l|c|c|c|}
\hline
        & Sogou-08 & Sogou-11 & Yahoo-10\\
\hline
$\tau_K(d,\tau)$ & $0.1273$ & $0.0914$ & $0.0878$ \\
$\rho_S(d,\tau)$ & $0.1721$ & $0.1224$ & $0.1175$ \\
\hline
\end{tabular}
\caption{Correlations between step-length and waiting time for the different data sets. The positive correlation coefficients
indicate that for human online search processes spatial and temporal activities are not independent, with
stronger correlations emerging for Sogou-08 than for Sogou-11 and Yahoo-10.}
\label{d_t_corr}
\end{table}

Table IV shows our results for the correlations between step-length $d$ and waiting time $\tau$.
Both Kendall's tau and Spearman's rho provide correlation coefficients close to or larger than 0.10,
indicating a weak positive correlation between step-length and waiting time. 
We also note that the correlations for Sogou-08 are larger than for Sogou-11 and Yahoo-10, illustrating again
that different mechanisms underlie the different click-through logs.
The positive correlation indicates that
the assumption of independence between spatial and temporal activities, valid for both Brownian motion
and L\'{e}vy flights, does not hold in a strict sense for human online search processes.

\begin{table}[]
\centering
\begin{tabular}{|c|c|c|}
\hline
correlation & \multicolumn{2}{c|}{Sogou-08}                                                                                                                                                              \\ \hline
$m$         & $\tau_K\left(\{d_i\},\{d_{i+m}\}\right)$ & $\rho_s\left(\{d_i\},\{d_{i+m}\}\right)$   \\ \hline
1                & $0.2763$                                & $0.3511$                               \\ \hline 
2                & $0.2761$                                & $0.3496$                               \\ \hline 
3                & $0.2653$                                & $0.3358$                               \\ \hline 
4                & $0.2608$                                & $0.3297$                               \\ \hline 
5                & $0.2568$                                & $0.3243$                               \\ \hline
6                & $0.2519$                                & $0.3176$                               \\ \hline
7                & $0.2509$                                & $0.3157$                               \\ \hline
8                & $0.2506$                                & $0.3149$                               \\ \hline 
9                & $0.2476$                                & $0.3105$                               \\ \hline 
10               & $0.2477$                                & $0.3102$                               \\ \hline 
20               & $0.2433$                                & $0.3001$                               \\ \hline 
30               & $0.2343$                                & $0.2852$                               \\ \hline
40               & $0.2321$                                & $0.2784$                               \\ \hline 
50               & $0.2227$                                & $0.2638$                               \\ \hline
\end{tabular}
\caption{Correlations between successive displacements. The positive correlation coefficients indicate 
the presence of long-term memory effects in human online searches. We only calculated correlations for $\{d_i\}$ with $i\ge 11$.}
\label{d_corr}
\end{table}

\begin{figure} [h]
\includegraphics[width=0.70\columnwidth,clip=true]{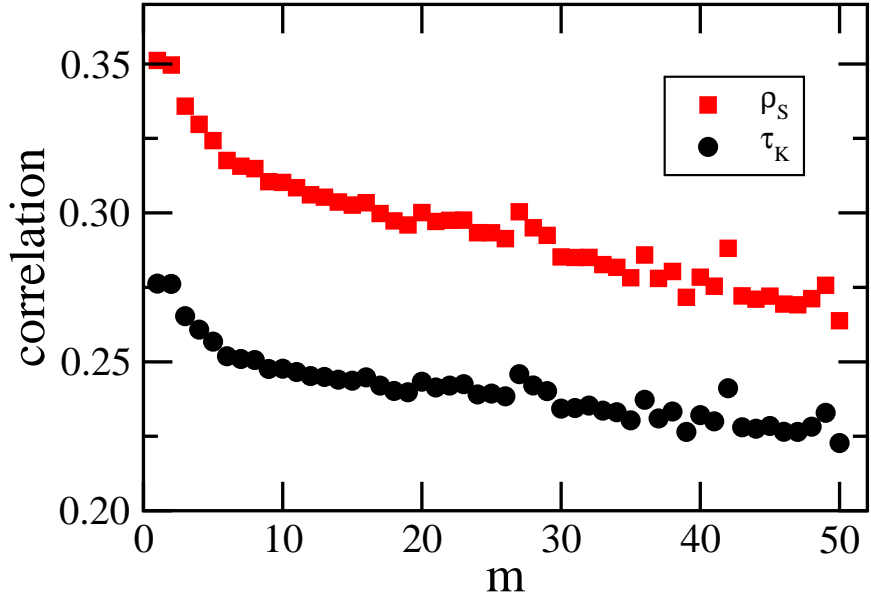}
\caption{\label{fig12}
(Color online) $\tau_K$ and $\rho_S$ as a function of $m$ for Sogou-08, see main text. These correlations decrease
for increasing $m$ but remain positive even for large $m$, revealing the presence of long-term memory effects in
online human searches.
}
\end{figure}

We can also check whether a process is memory-less or not. In order to do so we define for every step-length series 
$\left\{ d_i \right\} = \left\{d_1, d_2, \cdots \right\}$
sets $\left\{ d_{i+m} \right\} = \left\{d_{1+m}, d_{2+m} , \cdots \right\}$ with $m > 0$
and calculate the correlation coefficient $\tau_K(\left\{ d_i \right\},\left\{ d_{i+m} \right\})$ and
$\rho_S(\left\{ d_i \right\},\left\{ d_{i+m} \right\})$. For a completely memory-less process these
two coefficients should be zero. 
As we have previously seen that $S_d(n)$ exhibits a plateau, characteristic of 
a stationary process, only for Sogou-08 and $n \geq 11$, we performed this analysis only for Sogou-08 and series
$\{d_i\}$ with $d_i \geq 11$. The results shown in Table V and Fig. \ref{fig12} reveal positive correlations even for large values of $m$.
Long-term memory effects therefore permeate online human searches.

\section{Discussion and conclusion}
Online activities are an integral part of our daily lives. In many cases these activities involve search queries
submitted to one of the search engines. In this paper we propose to view the exploration of the results (i.e. links)
provided by the search engine as a foraging process on a semi-infinite line where the rank of a link corresponds
to a coordinate on that line. Using a variety of space- and time-dependent quantities we investigate three different publicly
available click-through logs.

Our study reveals a sharp contrast between the oldest log and the two logs obtained more recently. For the two newer
sets almost all queries can be understood as local searches restricted to a single page. These local searches have many
characteristics of a Brownian random walk, 
as for example an exponential decay of the step-length distribution
or a mean-square displacement that increases almost linearly with the clicking order. However, there are also marked
differences, as a power-law behavior of the waiting time distribution or the directionality of the jumps along the semi-infinite
line. A very interesting behavior emerges for the oldest data set, where both local searches as well as long-range 
power-law distributed jumps are found. This behavior is reminiscent of intermittent processes with 
L\'{e}vy strategies that have been proposed to describe searches with hidden targets. Interestingly, the waiting time
distributions, i.e. the distributions of the times elapsed between successive clicks, 
are very similar for all three click-through logs and reveal a power-law behavior with an exponential
cut-off as that encountered in other areas
of human activities.

The different properties of the different data sets point to the evolution of the search engines over the years.
Until recently search engines were of limited efficiency and as a result a sizeable number of queries ended
up with the user jumping from one page with links to another. 
These searches are therefore best characterized as a combination of local explorations and power law distributed
relocations. The more recent logs reveal
an increased efficiency of the search engines where the overwhelming part of the queries yield only local
searches on the first page of results. These local searches have some characteristics of Brownian motion.

The online search
behavior clearly changes as a function of the efficiency of a search engine. Whereas efficient search engines
result in an overwhelmingly large number of local searches, earlier engines prompted searches where local explorations
and long-range, power law distributed, relocations processes are combined. This scenario shows many common features
with those of intermittent search processes which have been proposed as search strategies for finding hidden resources. 

The description of online searches as foraging processes yields some interesting insights, but our results
also reveal notable deviations from the simple models used to describe foraging processes.
It remains a challenge to come up with a more realistic foraging model that is capable of reproducing
the results we obtain from our analysis of the different click-through logs.

\begin{acknowledgments}
This work is supported in part by the US National
Science Foundation through grant DMR-1205309.
\end{acknowledgments}

\appendix
\section{Data description and preparation}

For our study we analyzed three different click-through data sets. In the following we briefly describe
these sets as well as how we prepared the data and used them in our study.

Sogou-08 and Sogou-11 refer to the data sets {\it Search Engine Click-through Log Version 2008}
and {\it Search Engine Click-through Log Version 2011} \cite{Sogou} which are parts of the 
``Sogou Lab Data." These two data sets provide millions of users' search queries and 
click-through activities on Sogou (www.sogou.com), which is one of the largest Chinese
search engines. Sogou-08 has been collected in June 2008, whereas Sogou-11 contains queries 
submitted from 12/30/2011 to 01/01/2012. For Sogou-08 respectively Sogou-11  we get $51,537,388$
respectively $43,545,440$ lines of record, each corresponding to an individual click
and consisting of the following components \cite{Sogou}:\newline

\begin{table}[!h]
\begin{tabular}{c|c|c|c|c|c}
Time of click &
User ID &
User query &
Ranking of clicked result &
Order of click &
URL of clicked result
\end{tabular}
\end{table}

\noindent
where ``click" refers to the event of the user clicking on a link on the search engine result pages,
whereas ``time" is the calendar time of the click event.

Yahoo (YahooL18 \cite{Yahoo}) is the data set {\it Annonymized Yahoo! Search Logs with
Relevance Judgments version 1.0} provided by Yahoo! Labs \cite{yahoolabs} as part of the Yahoo! 
Webscope program \cite{webscope} (``Approval for Access" granted October 7, 2013). 
It provides users' search click-through logs on Yahoo Search
(search.yahoo.com) collected in July 2010. Yahoo-10 contains $80,779,266$ lines, where each
line corresponds to an individual search and contains the following information 
(separated by ``\textbackslash t") \cite{Yahoo}:

\begin{table}[!h]
\begin{tabular}{c|c|c|c|c|c}
Query&
Cookie&
Timestamp&
List of URLs&
Number of ``clicks"&
List of time and click position/type pairs
\end{tabular}
\end{table}

\noindent
where the ``List of URLs" is the list of weblinks on the first result page.
``Click" refers to any type of click on the result pages, including clicking on a search result, 
clicking at the top of a result page (unspecified; this can include clicking on the ``also-try" button, 
on a spell correction suggestion, on an advertisement located at the top of the page, etc.), 
clicking at the bottom of a result page
(unspecified, this can include clicking on next page button, on the bottom ``also-try" button, 
on an advertisement located at the bottom of the page, etc.). For the ``time and click position/type pairs", the ``time" is the time 
(in seconds) since the beginning of the search. 
Since we do not know the specific activities done when
clicking at the top or the bottom of a result page, we ignore the clicks at the top of result pages
but assume that a click at the bottom of a result page is on the next page button.

From all these sets we removed entries with missing values for the first click as well as (for Sogou-08) unusual cases where a
click was on a link of rank 1000 or above. For entries with successive clicks on the same link, we only kept the 
first click and jumped to the next click on a different link. The total number of queries retained after this
procedure are listed in Table I.

After this data cleaning all clicks belonging to the same search were grouped together and the corresponding
(time, rank) pairs were calculated based on the order of the clicks. Time is the time passed in seconds since 
the click on the first result. In this way we end up for each search with a series of (time, rank) pairs

\begin{table}[!h]
\begin{tabular}{c|c|c|c|c}
$(t_1,r_1)$&
$(t_2,r_2)$&
$(t_3,r_3)$&
$(t_4,r_4)$&
$\cdots$
\end{tabular}
\end{table}

\noindent
with $t_1=0$. These series were then used as the starting point for our study.

\section{Maximum likelihood estimators}
Let us consider first the discrete power-law distribution
\begin{equation}
\displaystyle P(k) = \frac{k^{-\alpha}}{\zeta(\alpha, k_\text{min})}.
\end{equation}
For a given data set $\{k_i\}$, we have the likelihood
\begin{equation}
\displaystyle {\mathcal L}(\alpha|\mathbf{k}) = \prod\limits_{i=1}^n P(k_i) = \frac{\left(\prod\limits_{i=1}^{n} k_i \right)^{-\alpha}}{\zeta \left(\alpha, k_\text{min}\right)^n},
\end{equation}
where $n$ is the number of data points. The log-likelihood is then given by
\begin{equation}
\displaystyle \ln {\mathcal L}(\alpha|\mathbf{k}) = -\alpha \sum\limits_{i=1}^n \ln k_i - n \ln\zeta(\alpha, k_\text{min})~,
\end{equation}
and the maximum likelihood estimator (MLE) for $\alpha$ is obtained numerically from \cite{Bauke2007}
\begin{equation}
\displaystyle \hat{\alpha} = \arg\max_{\alpha} \ln {\mathcal {L}}(\alpha|\mathbf{k})=\arg\max_{\alpha} \left( -\alpha \sum\limits_{i=1}^n \ln k_i - n \ln\zeta(\alpha, k_\text{min}) \right)~.
\end{equation}
We used the L-BFGS-B method for parameter optimization,
see for example Ref. \cite{Zhu1997}.


For the ``shifted'' geometric distribution
\begin{equation}
\displaystyle P(k) = p(1-p)^{k-k_{\rm{min}}}, \quad k \geq k_{\rm{min}}
\end{equation}
the likelihood is given by
\begin{equation}
\displaystyle {\mathcal L}(p|\mathbf{k}) = \prod\limits_{i=1}^n P(k_i) = (1-p)^{\sum\limits_{i=1}^n k_i - n k_{\rm{min}}} p^n,
\end{equation}
where $n$ is again the size of the data. 
The maximum likelihood estimator for $p$ is
\begin{equation}
\displaystyle \hat{p} = \frac{n}{\sum\limits_{i=1}^n k_i - n k_{\rm{min}} + n} = 
\frac{1}{\bar{k}-(k_{\rm{min}}-1)}
\end{equation}
where $\bar{k}$ is the mean of $k_i$'s. Finally, for the exponential form
\begin{equation}
\displaystyle P(k) = \left( 1-e^{-\lambda} \right) e^{-\lambda(k-k_\text{min})},
\end{equation}
we obtain 
\begin{equation}
\displaystyle \hat{\lambda} = -\ln(1-\hat{p})= -\ln\left( 1- \frac{1}{\bar{k}-(k_{\rm{min}}-1)} \right),
\end{equation}
when $\bar{k}> k_\text{min}$, since $\lambda = - \ln (1-p)$ and MLE is invariant to this transformation.

The MLE for parameters in the other distributions are obtained in similar ways. For the Yule-Simon distribution
the MLE for the shape parameter $\alpha$ is
\begin{equation} \label{alpha-yule-simon}
\displaystyle \hat{\alpha} = \arg\max_{\alpha} \left( n \ln(\alpha-1) +n \ln\Gamma\left( k_\text{min}+\alpha+1\right) - 
\sum_{i=1}^n \ln\Gamma\left( k_i + \alpha \right) \right),
\end{equation}
whereas for the conditional Poisson distribution the MLE for $\mu$ is
\begin{equation} \label{conditinoal-poisson}
\displaystyle \hat{\mu} = \arg\max_{\mu} \left( -n\mu -n\ln\left( 1- F_\mu(k_\text{min}-1 )\right) + \sum_{i=1}^n k_i \ln\mu \right),
\end{equation}
where $F_\mu(\cdot)$ is the cumulative distribution function of a Poisson distribution with rate parameter $\mu$.

Finally for the distributions with more than one parameter, one has:\\
power-law with exponential cut-off
\begin{equation} \label{mle-power-law-with-exponential-cutoff}
\displaystyle \lbrace \hat{\alpha}, \hat{\lambda} \rbrace = \arg\max_{\alpha, \lambda} \left( -n \ln\left( Li_\alpha(e^{-\lambda})-\sum_{i=1}^{k_{\text{min}-1}}i^{-\alpha} e^{-\lambda i} \right) -\alpha \sum_{i=1}^n \ln k_i -\lambda \sum_{i=1}^n k_i \right);
\end{equation}
\noindent
discrete log-normal
\begin{equation} \label{mle-discrete-lognormal}
\displaystyle \lbrace  \hat{\mu}, \hat{\sigma} \rbrace = \arg\max_{\mu, \sigma} \left(\sum_{i=1}^n\ln\left( \Phi \left(\frac{\ln(k_i+1)-\mu}{\sigma}\right)-\Phi\left(\frac{\ln(k_i)-\mu}{\sigma} \right) \right) - n \ln\left(1-\Phi \left(\frac{\ln(k_\text{min})-\mu}{\sigma} \right) \right) \right);
\end{equation}
\noindent
pairwise power law
\begin{equation}\label{mle-pairwise-powerlaw}
\displaystyle \lbrace \hat{\alpha}, \hat{\beta}, \hat{k}_\text{trans}\rbrace = \arg\max_{\alpha, \beta, k_\text{trans}}  \left( n \ln C - \alpha \sum_{i=1}^n \ln k_i -(\beta-\alpha) \sum_{k_i \ge \lceil k_\text{trans} \rceil} \left( \ln k_i- \ln k_\text{trans} \right) \right).
\end{equation}

\section{Correlation coefficients}
%

\subsection{Kendall's tau}
Kendall's tau provides a measure of rank correlation. Assuming a set of observations $(x_1,y_1)$, $(x_2,y_2)$, $\cdots$
of two joint random variables $x$ and $y$ (step-length and waiting time, for example), Kendall's tau compares
the number $P$ of concordant pairs with the number $Q$ of discordant pairs:
\begin{equation}
\tau_K(x,y) = \frac{P - Q}{P + Q}~.
\end{equation}
Two pairs $(x_l,y_l)$ and $(x_m,y_m)$ are concordant if (1) $x_l < x_m$ and $y_l < y_m$ or (2) 
$x_l > x_m$ and $y_l > y_m$. If, however, $x_l < x_m$ and $y_l > y_m$ or $x_l > x_m$ and $y_l < y_m$, then they
are discordant. In the case of data with tied ranks (which is the case for our series), Kendall's tau can be
calculated as
\begin{equation}
\tau_K(x,y) = \frac{\displaystyle \sum\limits_{i<j} \rm{sgn} \left[(x_i-x_j)(y_i-y_j) \right]}{\displaystyle \sqrt{\frac{1}{2}
n(n-1) - U} \sqrt{\frac{1}{2}n(n-1) - V}}
\end{equation}
where $\rm{sgn}$ is the signum function, whereas $U$ and $V$ are the numbers of $x$-tied pairs and $y$-tied pairs.

\subsection{Spearman's rho}
Spearman's rho is Pearson's correlation coefficient between ranked variables. Denoting by $u_i$ the rank of $x_i$
and by $v_i$ the rank of $y_i$, then Spearman's rho can be expresses as:
\begin{equation}
\displaystyle \rho_S(x,y) = \displaystyle \frac{\displaystyle \sum\limits_{i=1}^{n} (u_i-\bar{u})(v_i-\bar{v})}{\sqrt{\displaystyle \sum\limits_{i=1}^{n} (u_i-\bar{u})^2} \sqrt{\displaystyle \sum\limits_{i=1}^{n} (v_i-\bar{v})^2}}
\end{equation}
where $\bar{u} = \displaystyle \frac{1}{n} \sum\limits_{i=1}^{n} u_i$ and
$\bar{v} = \displaystyle \frac{1}{n} \sum\limits_{i=1}^{n} v_i$.
While Spearman's rho and Kendall's tau usually yield different numbers, they work the same way.


\begin{thebibliography}{99}
\bibitem{footnote1} The waiting time consists of two parts: the time for viewing the previously 
clicked result and the time needed to select the next link on the search result pages. Our
data do not allow us to distinguish between these two contributions.
\bibitem{footnote2} Whereas for Sogou-08 and Sogou-11 the logs are divided into different days, yielding a
cut-off of 86,400 s for the waiting time, for Yahoo-10 the longest allowed time for a search is one hour.
These cut-offs are readily seen in Fig. \ref{fig7}.
\end{thebibliography}
\end{document}